\documentclass[12pt]{article}
\usepackage{amsmath,epsfig}

\DeclareMathAlphabet\mathfrak{U}{euf}{m}{n}
\DeclareFontFamily{U}{euf}{}
\DeclareFontShape{U}{euf}{m}{n}{
  <5> <6> <7> <8> <9> gen * eufm
  <10> <10.95> <12> <14.4> <17.28> <20.74> <24.88> eufm10
  }{}
\DeclareMathAlphabet\mathbb  {U}{msb}{m}{n}
\DeclareFontFamily{U}{msb}{}
\DeclareFontShape{U}{msb}{m}{n}{
  <5> <6> <7> <8> <9> gen * msbm
  <10> <10.95> <12> <14.4> <17.28> <20.74> <24.88> msbm10
  }{}

\DeclareSymbolFont{AMSa}{U}{msa}{m}{n}
\DeclareMathSymbol{\gtrless}  {\mathrel}{AMSa}{"3F}
\DeclareMathSymbol{\gtreqless}{\mathrel}{AMSa}{"52}

\newtheorem{dfn}{Definition}
\def\dslash{\partial \kern-0.52em \raisebox{0.2ex}{/} \kern 0.1em}
\def\one{\mathsf{1}}
\allowdisplaybreaks[4]

\makeatletter
\def\section{\@startsection {section}{1}{\z@}{-3.25ex plus -1ex minus-.2ex}{
	     1.5ex plus.2ex}{\reset@font\normalsize\bfseries}}

\def\thebibliography#1{\section*{\refname}\list
  {\@biblabel{\theenumiv}}{\settowidth\labelwidth{\@biblabel{#1}}%
    \leftmargin\labelwidth
    \advance\leftmargin\labelsep
    \footnotesize \parsep=0pt \itemsep=0pt
    \usecounter{enumiv}%
    \let\p@enumiv\@empty
    \def\theenumiv{\arabic{enumiv}}}%
    \def\newblock{\hskip .11em plus.33em minus.07em}%
    \sloppy\clubpenalty4000\widowpenalty4000
    \sfcode`\.=1000\relax}

\def\ps@hep{\def\@oddhead{\hfil\texttt{hep-th/9712183}}\let\@evenhead\@oddhead
	\def\@oddfoot{\hfil\thepage\hfil}\let\@evenfoot\@oddfoot}
\makeatother

\begin{document}

\thispagestyle{hep}

\begin{center}
{\renewcommand{\thefootnote}{\fnsymbol{footnote}}
{\Large\bfseries Gyros as geometry of the standard model}
\\[3ex]
\textsc{Raimar Wulkenhaar}\footnote{\tabcolsep 0pt\begin{tabular}[t]{ll} 
1998 address:~{} & Centre de Physique Th\'eorique, CNRS Luminy, Case 907, \\ &
		   13288 Marseille Cedex 9, France \end{tabular}} 
\\[1ex]
{\small{\itshape Institut f\"ur Theoretische Physik, Universit\"at Leipzig\\
		Augustusplatz 10/11, D-04109 Leipzig, Germany}\\[0.5ex]
		e-mail: \texttt{raimar.wulkenhaar@itp.uni-leipzig.de}\\[1ex]
		December 18, 1997}
}\setcounter{footnote}{1}
\end{center}

\vskip 4ex

\begin{abstract}
We investigate the (noncommutative) geometry defined by the standard model, 
which turns out to be of Kaluza--Klein type. We find that spacetime points are 
replaced by extended two-dimensional objects which resemble the surface of a 
gyro. Their size is of the order of the inverse top quark mass.
\end{abstract}

\section{Introduction}

Gel'fand and Na\u{\i}mark realized \cite{gn} that a unital commutative 
$C^*$-algebra is essentially the same thing as a compact topological Hausdorff 
space. In the sequel mathematicians have dropped commutativity and considered 
noncommutative $C^*$-algebras as something like noncommutative topological 
spaces. Some highlights of this program are algebraic $K$-theory \cite{at}, 
cyclic cohomology \cite{cf,ac} and quantum groups \cite{dr,wo}. Physicists 
however are confronted with measurements, that is the assignment of a set of 
real numbers to the system under consideration. These real numbers constitute 
a metric space or geometry. Although topology has important applications to 
physics, geometry is indispensable. Therefore, Connes' assignment 
\cite{cf,ac,acr,cl} of metric properties to noncommutative topological spaces 
is of paramount importance for physics. 

Connes' discovery was that geometry is encoded in the interplay between a 
$*$-algebra $\mathcal{A}$ and some sort of Dirac operator $D$, both acting on 
a Hilbert space $\mathcal{H}$. The collection $(\mathcal{A},\mathcal{H},D)$ of 
these data is called spectral triple. Connes' distance definition, applied to 
the spectral triple (smooth functions on spin manifold $M$, Dirac operator of 
the spin connection, square integrable bispinors), recovers precisely 
\cite{cf,ac,cl} the geodesic distance on $M$. But the strength of Connes' 
definition is that it does not require the algebra $\mathcal{A}$ to be 
commutative. Moreover, it gives rise to an interesting geometry even on 
commutative algebras with noncommutative differential calculus, such as the 
famous two-point space \cite{cf,cl}. 

In spite of these opportunities, the interest of the physics community has 
moved more to the construction of differential calculi \cite{cl,kppw} 
associated to spectral triples and to physical models \cite{cl,cff,k1,vg} 
based on them. The formost achievement along this line is a reformulation of 
the standard model \cite{acr,iks,mgv,cc} in which Yang--Mills and Higgs fields 
are parts of one generalized gauge potential. This leads to a genuine 
unification of Yang--Mills and Higgs sectors of the standard model. However, 
I am not aware of an attempt to recover the metric structure associated to the 
standard model spectral triple, which is the concern of this paper.

\begin{figure}
\unitlength 1mm
\linethickness{1pt}
\hspace*{1cm}\begin{picture}(70,45)
\put(15,0){\epsfig{file=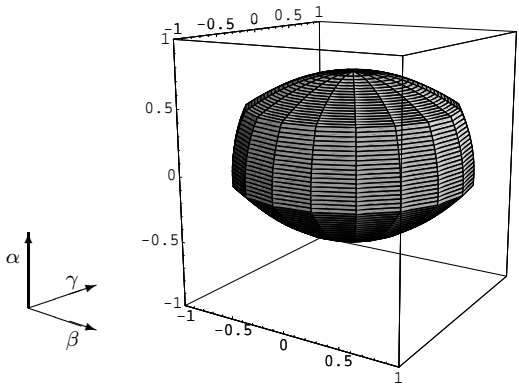,height=50mm,width=50mm}}
\put(0,10){\vector(3,-1){ 9}}
\put(0,10){\vector(3, 1){ 9}}
\put(0,10){\vector(0, 1){10}}
\put(-3,16){\scriptsize $\alpha$}
\put(5,5){\scriptsize $\beta$}
\put(5,13){\scriptsize $\gamma$}
\end{picture}\parbox[b]{48mm}{\caption{The gyro defined by 
$\delta=0.216$, $\epsilon=0.72$ \label{gyro}}}
\end{figure}

Technically, we do not strictly follow Connes' framework but employ the 
author's modification \cite{rw2} that uses Lie algebras instead of associative 
$*$-algebras (section \ref{fund}). The Lie algebraic framework has the 
advantage that every degree of freedom has a physical meaning, whereas in the 
$*$-algebra case one has additional parameters which at the very end are 
eliminated by skew-adjointness and unimodularity conditions. Our procedure  
implements neither real structures \cite{acr} nor all that sophisticated stuff 
like bivariant $K$-theory \cite{gk} and noncommutative Poincar\'e duality, 
which according to Connes \cite{ac,acr} are essential elements of 
noncommutative manifolds. We start with the pure matrix part of the standard 
model (section \ref{matrix}) and find that its geometry is a nine-parametric 
family (because there are nine massless Yang--Mills fields in the standard 
model) of infinitely distant two-dimensional objects. They are the surface of 
the unit ball whose polar regions are rotary-grinded to paraboloids, see figure 
\ref{gyro} for an example (the meaning of the coordinates 
$\alpha,\dots,\epsilon$ is explained in section \ref{matrix}). 
We call such an object a \emph{gyro}. The distance between points on the gyro 
equals $1/m_t$ times the Euclidean three-dimensional distance through the 
interior of the gyro, where $m_t$ is the mass of the top quark. 
The pure continuum case leads back to Riemannian geometry (section \ref{cont}). 

Thus, the geometry of the full standard model (section \ref{full}) is of 
Kaluza--Klein type \cite{ka,kl}. It is a nine-parametric family of infinitely 
distant worlds. Each world is six-dimensional (see also \cite{m}), four 
dimensions are our usual spacetime and the other two are compactified to a 
certain gyro. This means that we do not confirm Connes' conjecture of a 
multi-sheeted structure of the universe \cite{cf,ac}. It is true that the 
geometry of the standard model differs from four dimensional Riemannian 
geometry at energy scales of the order $m_t$. But on each world the geometry 
remains continuously connected and can be described completely in terms of 
standard (commutative) geometry. This is also in contrast with noncommutative 
Kaluza--Klein theories developed by Madore and Mourad (see \cite{mm} for a 
review and references therein), where the internal coordinates are generators 
of a noncommutative algebra. We show that the geometry of the matrix part of 
the standard model (which contains three massive Yang--Mills fields) is a 
deformation of the 2-sphere $S^2$. The spectral triple over the algebra 
$\mathbb{C} \oplus \mathbb{C}$ studied first by Connes and Lott \cite{cf,cl} 
gives rise to one massive Yang--Mills field. Therefore, its geometry is a 
deformed $S^0$, i.e.\ a pair of points. After taking spacetime into 
consideration, Connes and Lott thus obtained two copies of spacetime as 
geometry of this example. The possibility of endowing discrete spaces with 
geometry has been celebrated as a main achievement of noncommutative geometry. 
To my knowledge, one has widely believed that the discreteness of the 
$\mathbb{C} \oplus \mathbb{C}$-example is typical for matrix spectral triples. 
But this is not the case, as the present paper shows.

\section{Fundamentals}
\label{fund}

Physical reasons (the wish to describe other field theoretical models than the 
standard model) led us to replace the associative $*$-algebra in Connes' 
noncommutative geometry by a Lie algebra \cite{rw2}. Then, the spectral triple 
or $K$-cycle describing the initial data becomes an $L$-cycle:
\begin{dfn} 
An $L$--cycle $(\mathfrak{g},\mathcal{H},D,\pi,\Gamma)$ over a skew--adjoint 
Lie algebra $\mathfrak{g}$ is given by 
\begin{list}{}{\settowidth{\labelwidth}{iii)}
	       \leftmargin\labelwidth\addtolength{\leftmargin}{\labelsep}
	       \itemsep0pt\parsep0pt\parskip0pt\topsep0pt\partopsep0pt}

\item[\textup{i)}] an involutive representation $\pi$ of $\mathfrak{g}$ in the 
Lie algebra $\mathcal{B}(\mathcal{H})$ of bounded operators on a Hilbert 
space $\mathcal{H}$, i.e. $(\pi(a))^*=\pi(a^*) \equiv -\pi(a)$, for any 
$a \in \mathfrak{g}$,

\item[\textup{ii)}] a (possibly unbounded) selfadjoint operator $D$ on 
$\mathcal{H}$ with compact resolvent such that $[D,\pi(a)] \in 
\mathcal{B}(\mathcal{H})$,

\item[\textup{iii)}] a selfadjoint operator $\Gamma$ on $\mathcal{H}$, 
fulfilling $\Gamma^2 = \mathrm{id}_\mathcal{H}$, $\Gamma D + D \Gamma=0$ 
and $\Gamma \pi(a) - \pi(a)\Gamma=0$.
\end{list}
\label{lcc} 
\end{dfn} 

We recall \cite{rw2} the definition of a metric structure on $L$-cycles, 
obtained by a simple adaptation of Connes' proposal to our case:
\begin{dfn} 
Let $\mathrm{X}$ be the space of linear functionals $\chi$ on $\mathfrak{g}$ 
whose norm equals $1$, i.e.\ $\|\chi\| =\sup_{a \in \mathfrak{g}} \big( 
|\chi(a)| / \|\pi(a)\|\big)=1$. The distance $\mathrm{dist}(\chi_1,\chi_2)$ 
between $\chi_1,\chi_2 \in \mathrm{X}$ is given by 
\begin{equation}
\mathrm{dist}(\chi_1,\chi_2):=\sup_{a \in \mathfrak{g}} 
\{~|\chi_1(a) -\chi_2(a)|~:~~ \|\,[D, \pi(a)] \,\| \leq 1 ~\}~. 
\end{equation}
\end{dfn}
The concern of this paper is to show the usefulness of this definition 
by means of a commutative example (section \ref{commut}) and to investigate 
the metric structure of the standard model $L$-cycle.

\section{The commutative case}
\label{commut}

The Dirac $K$-cycle $(C^\infty(M),L^2(\mathcal{S}),\mathrm{i} \dslash)$ can be 
regarded as an $L$-cycle over the commutative Lie algebra 
$\mathrm{i} C^\infty(M)$ as well. Here, $M$ is the Euclidean spacetime 
(4-dimensional compact Riemannian spin manifold), $C^\infty(M)$ denotes the 
algebra of real-valued smooth functions over $M$, $L^2(\mathcal{S})$ is the 
Hilbert space of square integrable sections of the spinor bundle $\mathcal{S}$ 
over $M$ and $\mathrm{i} \dslash =\mathrm{i} \gamma^\mu \partial_\mu$ is the 
Dirac operator of the spin connection. From Connes' discovery \cite{cf,ac,cl} 
that the Dirac $K$-cycle gives rise to Riemannian geometry on $M$ we expect 
that this is also true for the $L$-cycle. 

We compute the distance between those linear functionals $\chi_p,\chi_q$ on 
$\mathrm{i} C^\infty(M)$ which are even characters determined by points of $M$, 
i.e.\ $\chi_p(\mathrm{i} f)=f(p)$, $\chi_q(\mathrm{i} f)=f(q)$, for 
$\mathrm{i} f \in \mathrm{i} C^\infty(M)$ and $p,q \in M$. 
Obviously, 
\[
\|\chi_p\|=\sup_{f\in C^\infty(M)} \frac{|f(p)|}{\|f\|}=
\sup_{f\in C^\infty(M)} \frac{|f(p)|}{\max_{q \in M} |f(q)|}=1~. 
\]
We have 
\begin{equation}
[D,\pi(f)] \equiv [\mathrm{i}\dslash,f]
=\mathrm{i} \gamma^\mu \frac{\partial f}{\partial x^\mu} 
=\mathrm{i} \gamma (dx^\mu) \frac{\partial f}{\partial x^\mu} 
=\mathrm{i} \gamma (d f)~,
\label{Df0}
\end{equation}
where $\gamma:\Lambda^1(M) \to \mathcal{B}(L^2(\mathcal{S}))$ is the Clifford 
representation of sections of the cotangent bundle. The Clifford representation 
fulfills $\gamma(\omega) \gamma(\omega)=g^{-1}(\omega,\omega) \one_4$, for 
$\omega \in \Lambda^1(M)$, where the bilinear form on sections of the cotangent 
bundle $g^{-1}:\Lambda^1(M) \times \Lambda^1(M) \to C^\infty(M)$ is the inverse 
of the metric $g$. The norm is obtained by optimization of differentiation 
along a curve $\mathcal{C}(s)$:
\begin{align}
\|[D,\pi(f)]\| &=\|\mathrm{i} \gamma (df)\|
=\sup_{\mathcal{C}}\left|\!\left|\gamma (ds) \dfrac{df}{ds}\right|\!\right|
=\sup_{\mathcal{C}}\left|\!\left|\gamma (ds) \gamma (ds) 
  \left|\dfrac{df}{ds}\right|^2 \right|\!\right|^{1/2} \notag \\
& = \sup_{\mathcal{C}}\left|\!\left| g^{-1}(ds,ds) \left|\dfrac{df}{ds}
\right|^2 \right|\!\right|^{1/2}
= \sup_{\mathcal{C},x} \left\{ \left|\frac{df}{ds}\right| \sqrt{g^{-1}(ds,ds)} 
\right\}~,  \label{Df}
\end{align}
where the supremum is taken over all points $x \in M$ and all 
curves $\mathcal{C}=\mathcal{C}(s)$ through $x$. 

We consider only curves where $s$ is the arc length satisfying 
$g^{-1}(ds,ds)=1$. This means that $|s(p)-s(q)|=\mathrm{length}(\widehat{pq})$ 
is the length of the curve connecting $p,q \in M$. We write the 
differentiation in \eqref{Df} as the limit of a difference quotient:
\[
\sup_{\mathcal{C},x} \left|\frac{df}{ds}\right| 
=\sup_{\mathcal{C},x} \lim_{p\to x,\;p \in \mathcal{C}} 
\left|\frac{f(x)-f(p)}{s(x)-s(p)}\right| 
=\sup_{\mathcal{C},x} \lim_{p\to x,\;p \in \mathcal{C}} 
\frac{|f(x)-f(p)|}{\mathrm{length}(\widehat{xp})}~.
\]
This value is maximized if $\mathcal{C}$ is a geodesics connecting $x$ and $p$, 
and instead of varying $\mathcal{C}$ we can equivalently vary the point $p$ 
that defines the geodesics $\mathcal{C}$. Moreover, the following consideration
\[
|f(x)-f(p)|= \left|\int_{p}^{x} \frac{df}{ds} ds\right| 
\leq \left| \int_{p}^{x} \!\!ds\right| \;\max_{q \in \widehat{xp}} 
\left|\frac{df}{ds}\right| = \mathrm{dist}(x,p)\,
\max_{q \in \widehat{xp}}  \left|\frac{df}{ds}\right| 
\]
shows that 
\[
\sup_x \frac{|f(x)-f(p)|}{\mathrm{dist}(x,p)} \leq 
\sup_x \lim_{p\to x} \frac{|f(x)-f(p)|}{\mathrm{dist}(p,q)} ~.
\]
This gives
\begin{equation}
\|[D,\pi(f)]\|=\sup_{x \neq p} \frac{|f(x)-f(p)|}{\mathrm{dist}(x,p)}~,
\label{Df1}
\end{equation}
as stated in \cite{cf,ac}. Therefore, $\|[D,\pi(f)]\| \leq 1$ implies 
$|f(p)-f(q)| \leq \mathrm{dist}(p,q)$ for all functions $f$ under 
consideration and all points $p,q$, which means $\mathrm{dist}(\chi_p,\chi_q) 
\leq \mathrm{dist}(p,q)$. Taking in particular the distance function itself, 
$f_p(q)=\mathrm{dist}(p,q)$, one has $|f_p(q_1)-f_p(q_2)| 
=|\mathrm{dist}(p,q_1)-\mathrm{dist}(p,q_2)|\leq \mathrm{dist}(q_1,q_2)$ due to 
the triangle inequality, therefore, $\|[D,\pi(f_p)]\|\leq 1$ on one hand and 
$|f_p(p)-f_p(q)|=\mathrm{dist}(p,q)$ on the other hand. This means 
$\mathrm{dist}(\chi_p,\chi_q) = \mathrm{dist}(p,q)$. 

\section{The matrix part of the standard model}
\label{matrix}

The $L$-cycle of the matrix part of the standard model is the direct 
transcription of the physical situation and was already presented in 
\cite{rw3}. The Hilbert space is $\mathcal{H}=\mathbb{C}^{48}$ if we include 
right neutrinos. The Lie algebra is of course 
{\allowdisplaybreaks[1]
\begin{align}
\mathfrak{g} &=\mathrm{su(3)} \oplus \mathrm{su(2)} \oplus \mathrm{u}(1) 
\notag \\ 
& \ni \{\boldsymbol{g},\boldsymbol{a},e\} \equiv 
\{\mathrm{i} ({\textstyle \sum_{j=1}^8}\, g^j \lambda^j)\,,\, \mathrm{i}(a 
\sigma^3{+}b\sigma^1{+}c\sigma^2)\,,\,e\}~,
\label{lie321}
\end{align}}
where $\lambda^j$ are the Gell-Mann matrices, $\sigma^k$ the Pauli matrices and 
$g^j,a,b,c,e \in \mathbb{R}$. This Lie algebra acts on $\mathcal{H}$ via the 
representation
\begin{align}
\pi(\boldsymbol{g},\boldsymbol{a},e) &=\mbox{\small$ \left( \begin{array}{cc}
\pi_\ell(\boldsymbol{g},\boldsymbol{a},e) & 0 \\
0 & \pi_q(\boldsymbol{g},\boldsymbol{a},e) \end{array} \right) $}\;, \notag \\
\pi_\ell(\boldsymbol{g},\boldsymbol{a},e) &= \mbox{\small$ 
\left( \begin{array}{cccc} 
\mathrm{i} (a-e) \otimes \one_3 & \mathrm{i} (b -\mathrm{i}c) \otimes \one_3 
& 0 & 0\\
\mathrm{i} (b +\mathrm{i} c) \otimes \one_3 & \mathrm{i} (-a-e) \otimes \one_3 
& 0 & 0\\
0 & 0 & 0_3 & 0 \\
0 & 0 & 0 & -2 \mathrm{i} e \otimes \one_3 \end{array} \right) $},
\label{pi}
\\
\pi_q(\boldsymbol{g},\boldsymbol{a},e) &=\! \mbox{\small$\left( 
\renewcommand{\arraycolsep}{0pt} \begin{array}{cccc} 
(\mathrm{i} (a{+}\tfrac{1}{3} e) \one_3 {+} \boldsymbol{g}) {\otimes} \one_3 & 
\mathrm{i} (b {-}\mathrm{i}c) \one_3 {\otimes} \one_3 & 0 & 0\\
\mathrm{i} (b {+}\mathrm{i} c) \one_3 \otimes \one_3 & 
(\mathrm{i} ({-}a{+}\tfrac{1}{3} e) \one_3 {+} \boldsymbol{g}) {\otimes} \one_3 
& 0 & 0\\
0 & 0 & \hspace*{-1.5em} (\tfrac{4}{3} \mathrm{i} e \one_3 {+} \boldsymbol{g}) 
{\otimes} \one_3 \!\!\!\!
& 0 \\
0 & 0 & 0 & \!\!\!({-}\tfrac{2}{3} \mathrm{i} e \one_3 
{+} \boldsymbol{g}) {\otimes} 
\one_3	\end{array} \right) $}. \notag
\end{align}
The generalized Dirac operator is the Yukawa operator
\begin{align}
Y &= \mbox{\small$ \left( \begin{array}{cc} Y_\ell & 0 \\ 0 & Y_q 
\end{array} \right) $}\;, 
\label{Y} \\
Y_\ell &= \mbox{\small$ \left(\renewcommand{\arraycolsep}{2pt} 
\begin{array}{cccc} 
0 & 0 & \mathcal{M}_\nu & 0 \\
0 & 0 & 0 & \mathcal{M}_e \\
\mathcal{M}_\nu^* & 0 & 0 & 0 \\
0 & \mathcal{M}_e^* & 0 & 0 \end{array} \right) $}, \quad
Y_q= \mbox{\small$ \left( \renewcommand{\arraycolsep}{0pt}\begin{array}{cccc} 
0 & 0 & \one_3 \otimes \mathcal{M}_u & 0 \\
0 & 0 & 0 & \one_3 \otimes \mathcal{M}_d \\
\one_3 \otimes \mathcal{M}_u^* & 0 & 0 & 0 \\
0 & \one_3 \otimes \mathcal{M}_d^* & 0 & 0 \end{array} \right) $},  \notag
\end{align}
where $\mathcal{M}_{e,\nu,u,d}$ are $3\times 3$-mass matrices of the fermions.

The first and most difficult part is to compute the norm of functionals 
$\chi$ on $\mathfrak{g}$. The computation consists of two steps: that of 
the norm of $\pi(\boldsymbol{g},\boldsymbol{a},e)$ and that of the extrema of 
$\chi(\boldsymbol{g},\boldsymbol{a},e)/\|\pi(\boldsymbol{g},
\boldsymbol{a},e)\|$. Let $\mathrm{i} g_i$, $g_1\leq g_2 \leq g_3$, 
be the eigenvalues of $\boldsymbol{g} \in \mathrm{su(3)}$. 
As $\boldsymbol{g}$ is tracefree, there are only the two possibilities
$g_1 \leq g_2 \leq 0 \leq g_3{=}|g_2|{+}|g_1|$ and 
$g_1{=}{-}|g_2|{-}|g_3| \leq 0 \leq g_2 \leq g_3$. It suffices to study the 
first case, because the second case goes into the first one by inversion 
$\{\boldsymbol{g},\boldsymbol{a},e\} \mapsto 
\{-\boldsymbol{g},-\boldsymbol{a},-e\}$. Thus, we have
\begin{equation}
g_1 \leq g_2 \leq 0 \leq |g_2| \leq \tfrac{1}{2} g_3 \leq |g_1| \leq 
g_3{=}|g_1|{+}|g_2|~.
\label{g}
\end{equation}
Denoting $\|\boldsymbol{a}\|:=\sqrt{a^2+b^2+c^2}$, the eigenvalues of 
$\pi(\boldsymbol{g},\boldsymbol{a},e)$ are $\mathrm{i}$ times
\begin{gather*}
-e+\|\boldsymbol{a}\|~,\quad -e-\|\boldsymbol{a}\|~,\quad -2e~, \\
\tfrac{1}{3} e +\|\boldsymbol{a}\| + g_i~,\quad 
\tfrac{1}{3} e -\|\boldsymbol{a}\| + g_i~,\quad 
\tfrac{4}{3} e + g_i~,\quad 
-\tfrac{2}{3} e + g_i~,
\end{gather*}
so that the norm (=absolute value of the largest eigenvalue) becomes 
\begin{equation}
\|\pi(\boldsymbol{g},\boldsymbol{a},e)\| = 
\max\big(|e|{+}\|\boldsymbol{a}\|\,,\, 2|e|\,,\, 
|\tfrac{1}{3} e {+} g_i| {+} \|\boldsymbol{a}\| \,,\,
|\tfrac{1}{3} e {+} g_i| {+} |e| \big)~. 
\label{max}
\end{equation}

For the further evaluation we draw the `left' graphs 
$y=|e|{+}\|\boldsymbol{a}\|$, $y=|\tfrac{1}{3} e {+} g_i| 
{+} \|\boldsymbol{a}\|$ and the `right' graphs
$y=2|e|$, $y=|\frac{1}{3} e {+} g_i| {+} |e|$ into a $e$-$y$-diagram. 
The partial norm functions of the `left' and `right' graphs are
\begin{equation}
{\renewcommand{\arraystretch}{1.4}
\begin{array}{ccc}
\mbox{interval} & \mbox{right norm} & \mbox{left norm} \\ \hline
e \leq -\tfrac{3}{2} |g_1| & -2 e & {-} e {+} \|\boldsymbol{a}\| 
\\
-\tfrac{3}{2} |g_1| \leq e \leq {-} \tfrac{3}{2}(g_3{-}|g_1|) & 
|g_1|{-}\tfrac{4}{3} e & |g_1|{-}\tfrac{1}{3} e {+} \|\boldsymbol{a}\| 
\\
{-} \tfrac{3}{2}(g_3{-}|g_1|) \leq e \leq 0 & g_3{-}\tfrac{2}{3} e 
& g_3{+}\tfrac{1}{3} e {+} \|\boldsymbol{a}\| 
\\
0 \leq e \leq \tfrac{3}{2} g_3 & g_3{+}\tfrac{4}{3} e 
& g_3{+}\tfrac{1}{3} e {+} \|\boldsymbol{a}\| 
\\
\tfrac{3}{2} g_3 \leq e & 2 e & e {+} \|\boldsymbol{a}\|
\end{array}}
\label{lrc}
\end{equation}
This table tells us that the left and right norms are continuous and piecewise 
linear with corners at 
\begin{equation}
e \in \big\{ \;- \tfrac{3}{2} |g_1| \,,\; -\tfrac{3}{2}(g_3-|g_1|)\,,\;0\,,\; 
\tfrac{3}{2} g_3 \;\big\}~.
\label{pointe}
\end{equation}
The total norm is the maximum of both partial norms and varies as we vary 
$\|\boldsymbol{a}\|$. The topology of the total norm function changes 
at those values of $\|\boldsymbol{a}\|$ where a corner of the left norm passes 
a corner of the right norm. These values are
\begin{equation}
\|\boldsymbol{a}\| \in \big\{ \;0\,,\; \tfrac{3}{2}(g_3-|g_1|)\,,\;
\tfrac{3}{2} |g_1| \,,\; \tfrac{3}{2} g_3 \;\big\}~.
\label{pointa}
\end{equation}

We consider functionals on $\mathrm{su(3)} \oplus \mathrm{su(2)} \oplus 
\mathrm{u(1)}$ of the form
\[
\chi_{\boldsymbol{\epsilon},\boldsymbol{\alpha},\delta}
(\boldsymbol{g},\boldsymbol{a},e)
:= \alpha a + \beta b + \gamma c + (2 \delta -\alpha) e 
+ {\textstyle \sum_{j=1}^8 \epsilon^j g^j~,}
\]
see \eqref{lie321} for the notation. In order to avoid the discussion of the 
eight parameters $\epsilon^j$ we adopt the following simplification. 
The norm of $\boldsymbol{\epsilon}$ is 
\[
\|\boldsymbol{\epsilon}\|=\sup_{\boldsymbol{g} \in \mathrm{su(3)}} 
{\textstyle |\sum_{j=1}^8 \epsilon^j g^j| \,/\, \|\boldsymbol{g}\|}~,
\]
where $\boldsymbol{g}$ is represented in the standard matrix representation of 
$\mathrm{su(3)}$. We assume $\boldsymbol{\epsilon}$ to be such that there is 
only one straight line through the origin in 
$\mathrm{su(3)} \cong \mathbb{R}^8$ on which the supremum of 
$|\sum_{j=1}^8 \epsilon^j g^j|/\|\boldsymbol{g}\|$ is 
attained. Under this condition we can define a sign of 
$\boldsymbol{\epsilon}$. We put $\mathrm{sign}(\boldsymbol{g})= 1$ if 
$|g_3| > |g_1|$ (which is our case), $\mathrm{sign}(\boldsymbol{g})= -1$ if 
$|g_3| < |g_1|$ and  
\begin{equation}
\textstyle \mathrm{sign} (\boldsymbol{\epsilon}) 
= \mathrm{sign} (\sum_{j=1}^8 \epsilon^j g^j) \;
\mathrm{sign}(\boldsymbol{g})~,\quad \mathrm{if}~~ 
|\sum_{j=1}^8 \epsilon^j g^j| = \|\boldsymbol{g}\|\;\|\boldsymbol{\epsilon}\|~.
\label{sign}
\end{equation}
This sign is constant on the maximal straight line (without $0$), and we have
$\mathrm{sign} (\boldsymbol{\epsilon})=-\mathrm{sign}(-\boldsymbol{\epsilon})$. 
We put $\epsilon=\mathrm{sign} (\boldsymbol{\epsilon})\: 
\|\boldsymbol{\epsilon}\|$.

Our goal is to examine the extrema of the functional $F$ on $\mathfrak{g}$ 
defined by $F(\boldsymbol{g},\boldsymbol{a},e)=\chi_{\boldsymbol{\epsilon},
\boldsymbol{\alpha},\delta}(\boldsymbol{g},\boldsymbol{a},e)\,/\, 
\|\pi(\boldsymbol{g},\boldsymbol{a},e)\|$. We have shown in \eqref{lrc} that 
$\|\pi(\boldsymbol{g},\boldsymbol{a},e)\|$ is a piecewise linear function of 
$\|\boldsymbol{a}\|,\;e,\;g_3,\;|g_1|$. Letting these parameters fixed, 
the extrema of $\chi_{\boldsymbol{\epsilon},\boldsymbol{\alpha},\delta}
(\boldsymbol{g},\boldsymbol{a},e)$ are taken at 
\[
\alpha a + \beta b + \gamma c = \pm \sqrt{\alpha^2+\beta^2+\gamma^2} 
\sqrt{a^2+b^2+c^2} \equiv \pm \|\boldsymbol{\alpha}\|\,\|\boldsymbol{a}\|
\]
(put $a=\pm (\alpha/ \|\boldsymbol{\alpha}\|)\,\|\boldsymbol{a}\|$,
$b=\pm (\beta/ \|\boldsymbol{\alpha}\|)\,\|\boldsymbol{a}\|$, 
$c=\pm (\gamma/ \|\boldsymbol{\alpha}\|)\,\|\boldsymbol{a}\|$). Moreover, we 
have $\sum_{j=1}^8 \epsilon^j g^j = \epsilon \|\boldsymbol{g}\|= \epsilon g_3$, 
see \eqref{sign}. Therefore, the candidates for 
$\chi_{\boldsymbol{\epsilon},\boldsymbol{\alpha},\delta}
(\boldsymbol{g},\boldsymbol{a},e)$ being extremal form the plane 
\[
\pm \|\boldsymbol{\alpha}\|\,\|\boldsymbol{a}\| + (2\delta {-} \alpha) e + 
\epsilon g_3~.
\]
But this means that $F$ is piecewise the quotient of first-order polynomials 
in the four parameters $\|\boldsymbol{a}\|,\;e,\;g_3,\;|g_1|$ and as such takes 
its extrema at the boundaries of the domain. In our case, these boundaries are 
the points specified in \eqref{pointe} and \eqref{pointa}. In order to find 
the extrema of $F$ it suffices to evaluate it at each combination of the points 
\eqref{pointe} and \eqref{pointa}. Moreover, we must take into account the 
various infinities of $e$ and $\|\boldsymbol{a}\|$ corresponding to case 
$\boldsymbol{g}=0$ and either $e=0$ or $\|\boldsymbol{a}\|=0$. The values of 
$F$ at all these points are given in table \ref{table}.
\begin{table}
{\small$\renewcommand{\arraystretch}{2.0}\begin{array}{rccl}
\mathrm{N}^{\mathrm{\underline{o}}} & e & \|\boldsymbol{a}\| & F 
\\ \hline
1 & {-} \tfrac{3}{2} |g_1| & 0 & 
\dfrac{{-}\tfrac{3}{2} (2 \delta{-}\alpha) |g_1| + \epsilon g_3}{ 3 |g_1|}
\\
2 & {-} \tfrac{3}{2} |g_1| & \tfrac{3}{2}(g_3 {-} |g_1|) & 
\dfrac{{-}\tfrac{3}{2} ((2 \delta{-}\alpha) {\pm} \|\boldsymbol{\alpha}\|) 
|g_1| + (\epsilon {\pm} \tfrac{3}{2} \|\boldsymbol{\alpha}\|) g_3 }{ 3 |g_1|}
\\
3 & {-} \tfrac{3}{2} |g_1| & \tfrac{3}{2}|g_1| & 
\dfrac{{-}\tfrac{3}{2} ((2 \delta{-}\alpha) {\mp} \|\boldsymbol{\alpha}\|) 
|g_1| + \epsilon g_3 }{3 |g_1|}
\\
4 & {-} \tfrac{3}{2} |g_1| & \tfrac{3}{2} g_3 & 
\dfrac{{-}\tfrac{3}{2} (2 \delta{-}\alpha) |g_1| 
+ (\epsilon {\pm} \tfrac{3}{2} \|\boldsymbol{\alpha}\|) g_3 }{ 
\tfrac{3}{2} (g_3{+}|g_1|)}
\\
5 & {-} \tfrac{3}{2} (g_3{-}|g_1|) & 0 & 
\dfrac{\tfrac{3}{2} (2 \delta{-}\alpha)|g_1|
+ (\epsilon {-}\tfrac{3}{2} (2 \delta{-}\alpha)) g_3}{ 2 g_3{-}|g_1|}
\\
6 & {-} \tfrac{3}{2} (g_3{-}|g_1|) & \tfrac{3}{2}(g_3{-}|g_1|) & 
\dfrac{ \tfrac{3}{2} ((2 \delta{-}\alpha) {\mp} \|\boldsymbol{\alpha}\|) 
|g_1| + (\epsilon {-} \tfrac{3}{2} ((2 \delta{-}\alpha) {\mp} 
\|\boldsymbol{\alpha}\|)) g_3 }{ 2 g_3 {-} |g_1|}
\\
7 & {-} \tfrac{3}{2} (g_3{-}|g_1|) & \tfrac{3}{2}|g_1| & 
\dfrac{\tfrac{3}{2} ((2 \delta{-}\alpha) {\pm} \|\boldsymbol{\alpha}\|) 
|g_1| + (\epsilon {-} \tfrac{3}{2}(2 \delta {-} \alpha)) g_3 }{
\tfrac{1}{2} g_3 {+} 2 |g_1|}
\\
8 & {-} \tfrac{3}{2} (g_3{-}|g_1|) & \tfrac{3}{2} g_3 & 
\dfrac{\tfrac{3}{2} (2 \delta{-}\alpha) |g_1| 
+ (\epsilon {-} \tfrac{3}{2}((2\delta {-} \alpha) 
{\mp} \|\boldsymbol{\alpha}\|)) g_3 }{ 2 g_3{+} \tfrac{1}{2} |g_1|}
\\
9 & 0 & 0 & \dfrac{\epsilon g_3}{g_3}
\\
10 & \tfrac{3}{2} g_3 & 0 & 
\dfrac{(\epsilon {+}\tfrac{3}{2} (2 \delta{-}\alpha)) g_3 }{ 3 g_3}
\\
11 & \tfrac{3}{2} g_3 & \tfrac{3}{2}(g_3{-}|g_1|) & 
\dfrac{{\mp} \tfrac{3}{2} \|\boldsymbol{\alpha}\| |g_1| 
+ (\epsilon {+} \tfrac{3}{2} ((2 \delta{-}\alpha) {\pm} 
\tfrac{3}{2} \|\boldsymbol{\alpha}\|)) g_3 }{ 3 g_3}
\\
12 & \tfrac{3}{2} g_3 & \tfrac{3}{2} |g_1| & 
\dfrac{{\pm} \tfrac{3}{2} \|\boldsymbol{\alpha}\| |g_1| 
+ (\epsilon {+} \tfrac{3}{2} (2 \delta{-}\alpha)) g_3 }{ 3 g_3}
\\
13 & \tfrac{3}{2} g_3 & \tfrac{3}{2} g_3 & 
\dfrac{(\epsilon {+} \tfrac{3}{2} ((2 \delta{-}\alpha) {\pm} 
\|\boldsymbol{\alpha}\|)) g_3 }{ 3 g_3} 
\\
14 & {\pm} M & 0 & \dfrac{{\pm}(2 \delta{-}\alpha) M}{2 M}
\\
15 & - M & M & \dfrac{({\pm}\|\boldsymbol{\alpha}\| {-} 
(2 \delta{-}\alpha)) M}{2 M}
\\
16 & M & M & \dfrac{({\pm}\|\boldsymbol{\alpha}\| {+} 
(2 \delta{-}\alpha)) M}{2 M}
\\
17 & 0 & M & \dfrac{{\pm}\|\boldsymbol{\alpha}\| M}{M}
\end{array}$}
\caption{The value of $F$ at the corner points}
\label{table}
\end{table}
In this table, we used that at $e=0$ there is a corner only if 
$\|\boldsymbol{a}\|=0$ and took the limit $M \to \infty$. The resulting 
function $F$ is still the quotient of first-order polynomials if it depends on 
$g_1$. The extrema are again attained at the boundaries $|g_1|=g_3$ 
(superscript $+$) and $|g_1|=\tfrac{1}{2} g_3$ (superscript $-$), see 
\eqref{g}. This gives the following absolute values:
\[
\renewcommand{\arraystretch}{1.5}
\begin{array}{ll|ll}
|F| & \mbox{$\mathrm{N}^{\mathrm{\underline{o}}}$ in Tab.\ \ref{table}} & 
|F| & \mbox{$\mathrm{N}^{\mathrm{\underline{o}}}$ in Tab.\ \ref{table}} 
\\ \hline
|\delta{-}\tfrac{1}{2} \alpha {-} \tfrac{1}{3} \epsilon| & 1^+,2^+
&
|\delta{-}\tfrac{1}{2} \alpha {-} \tfrac{2}{3} \epsilon| & 1^-,5^-
\\
|\delta{-}\tfrac{1}{2} \alpha {-} \tfrac{1}{3} \epsilon| {+} \tfrac{1}{2} 
\|\boldsymbol{\alpha}\| & 3^+,4^+
&
|\delta{-}\tfrac{1}{2} \alpha {-} \tfrac{2}{3} \epsilon| {+} \tfrac{1}{2} 
\|\boldsymbol{\alpha}\| 
& 2^-,3^-,6^-,7^-
\\
\tfrac{2}{3} |\delta{-}\tfrac{1}{2} \alpha {-} \tfrac{2}{3} \epsilon| 
{+} \tfrac{2}{3} \|\boldsymbol{\alpha}\| & 4^-,8^-
&
|\epsilon| & 5^+,6^+,9
\\
\tfrac{2}{5} |\epsilon| {+} \tfrac{3}{5} \|\boldsymbol{\alpha}\| & 7^+,8^+
&
|\delta{-}\tfrac{1}{2} \alpha {+} \tfrac{1}{3} \epsilon| & 10,11^+
\\
|\delta{-}\tfrac{1}{2} \alpha {+} \tfrac{1}{3} \epsilon| 
{+} \tfrac{1}{4} \|\boldsymbol{\alpha}\| & 11^-,12^-
&
|\delta{-}\tfrac{1}{2} \alpha {+} \tfrac{1}{3} \epsilon| 
{+} \tfrac{1}{2} \|\boldsymbol{\alpha}\| & 12^+,13
\\
|\delta{-}\tfrac{1}{2} \alpha| & 14
&
|\delta{-}\tfrac{1}{2} \alpha| {+} \tfrac{1}{2} \|\boldsymbol{\alpha}\| & 15,16
\\
\|\boldsymbol{\alpha}\| & 17
\end{array}
\]
After selecting the strongest constraints we find for $\|F\|=\max |F|$ 
\begin{equation}
\|F\| = \max\big( ~|\epsilon| \,,~ \|\boldsymbol{\alpha}\| \,,~
|\delta{-}\tfrac{1}{2} \alpha {-} \tfrac{1}{6} \epsilon| {+} \tfrac{1}{2} 
\|\boldsymbol{\alpha}\| {+} \tfrac{1}{2} |\epsilon| ~ \big)~.
\end{equation}
The requirement $\|F\|=\|\chi_{\boldsymbol{\epsilon},\boldsymbol{\alpha},
\delta}\|=1$ yields the following constraints:
\begin{equation}
|\epsilon|\leq 1 \,,~ \|\boldsymbol{\alpha}\| \leq 1 \,,~
|\delta{-}\tfrac{1}{6} \epsilon {-} \tfrac{1}{2} \alpha | 
+ \tfrac{1}{2} \|\boldsymbol{\alpha}\| \leq A\,,\quad 
A:=1 {-} \tfrac{1}{2} |\epsilon| \;,
\label{A}
\end{equation}
where at least one of these constraints must be an equality. 
The first conclusion is $|\delta{-}\tfrac{1}{6} \epsilon| \leq A$, which we 
satisfy by 
\begin{equation}
\delta= \tfrac{1}{6} \epsilon + A \cos \phi~,\quad 0 \leq \phi \leq \pi~.
\label{delta}
\end{equation}
This gives with $\|\boldsymbol{\alpha}\|^2 = \alpha^2+\beta^2+\gamma^2$ the 
equation 
\[
|A \cos \phi - \tfrac{1}{2} \alpha | + \tfrac{1}{2} 
\sqrt{\alpha^2+\beta^2+\gamma^2} = A~, 
\]
which leads (for the admissible range of $\alpha$ to be specified below) to
\begin{equation}
\beta^2+\gamma^2= 4 (A \pm A \cos \phi)^2 \mp 4 \alpha (A \pm A \cos \phi)~, 
\quad \mbox{for}~~\alpha \gtreqless 2 A \cos \phi~.
\end{equation}
What we obtain is thus a pair of paraboloids, which we can parametrize as 
follows:
\begin{equation}
\begin{split}
\alpha &= A(\cos \phi+ \cos \phi')~,\quad 0 \leq \phi' \leq \pi~,\\
\beta^2+\gamma^2 & = \mbox{\small$ \left\{ \begin{array}{c} 
4 A^2 (1+\cos \phi)(1-\cos \phi') \\
4 A^2 (1-\cos \phi)(1+\cos \phi') \end{array}\right.$} ~~ \mbox{for} ~~ 
\mbox{\small$ \begin{array}{c} \cos \phi \leq \cos \phi' \\ 
\cos \phi \geq \cos \phi' \end{array} $}
\end{split}
\label{alpha}
\end{equation}
However, we still have to take the condition $\|\boldsymbol{\alpha}\|\leq 1$ 
into account, which is the 2-sphere of radius $1$. This sphere will somewhere 
intersect the paraboloids \eqref{alpha} so that the total geometry is the 
composition of the sphere with one or two paraboloids replacing the sphere's 
polar regions. We call such an object a \emph{gyro}, which is the 
surface of a ball whose polar regions are rotary-grinded to paraboloids. 
Formulae \eqref{alpha} are equivalent to $\|\boldsymbol{\alpha}\| 
=A(2-|\cos \phi {-} \cos \phi'|)$, which determines the intersection 
parallels of latitude as $|\cos \phi - \cos \phi'| 
= \tfrac{2-2|\epsilon|}{2-|\epsilon|}$. Thus, for 
\begin{equation}
\left. \begin{array}{rcl}
\cos \phi' {-} \cos \phi &\!{\geq}\!& \tfrac{2-2|\epsilon|}{2-|\epsilon|}\\
|\cos \phi {-} \cos \phi'| &\!{\leq}\!& \tfrac{2-2|\epsilon|}{2-|\epsilon|}\\
\cos \phi {-} \cos \phi' &{\!\geq\!}& \tfrac{2-2|\epsilon|}{2-|\epsilon|}
\end{array} \right\}
\mbox{ we are on the } \left\{ \begin{array}{c}
\mbox{northern paraboloid} \\[0.3ex] \mbox{sphere}  \\[0.3ex] 
\mbox{southern paraboloid} 
\end{array} \right.
\label{phi}
\end{equation}

This situation is worth discussing. One has to choose $\epsilon \in [-1,1]$ 
and $\cos \phi \in [-1,1]$, which yields $\delta$ according to \eqref{delta} 
and \eqref{A}. Next, one chooses $\cos \phi'\in [-1,1]$ and determines from 
\eqref{phi} to which rotary body this value belongs. The height parameter 
$\alpha$ and in the paraboloid cases the radius $\sqrt{\beta^2+\gamma^2}$ are
obtained from \eqref{alpha}. In the spherical case we have of course 
$\sqrt{\beta^2+\gamma^2}=\sqrt{1-\alpha^2}$. Some special cases of these 
gyros are interesting (see figure \ref{fig}):
\begin{figure}
\unitlength 0.5mm
\linethickness{0.4pt}
\hspace*{2cm}\raisebox{5mm}{\begin{picture}(80,80)
\put(0,0){\line(1,0){80}}
\put(0,0){\line(0,1){80}}
\put(80,0){\line(0,1){80}}
\put(0,80){\line(1,0){80}}
\put(10,30){\line(0,1){30}}
\put(10,30){\line(3,-2){30}}
\put(10,60){\line(3,1){30}}
\put(40,10){\line(3,1){30}}
\put(40,70){\line(3,-2){30}}
\put(70,20){\line(0,1){30}}
\put(10,30){\line(3,1){60}}
\put(10,60){\line(3,-2){60}}
\put(40,-2){\line(0,1){4}}
\put(10,-2){\line(0,1){4}}
\put(70,-2){\line(0,1){4}}
\put(40,78){\line(0,1){4}}
\put(10,78){\line(0,1){4}}
\put(70,78){\line(0,1){4}}
\put(-2,10){\line(1,0){4}}
\put(-2,20){\line(1,0){4}}
\put(-2,30){\line(1,0){4}}
\put(-2,40){\line(1,0){4}}
\put(-2,50){\line(1,0){4}}
\put(-2,60){\line(1,0){4}}
\put(-2,70){\line(1,0){4}}
\put(78,10){\line(1,0){4}}
\put(78,20){\line(1,0){4}}
\put(78,30){\line(1,0){4}}
\put(78,40){\line(1,0){4}}
\put(78,50){\line(1,0){4}}
\put(78,60){\line(1,0){4}}
\put(78,70){\line(1,0){4}}
\put( 7,-9){\scriptsize -1}
\put(39,-9){\scriptsize  0}
\put(69,-9){\scriptsize  1}
\put( 7,85){\scriptsize -1}
\put(39,85){\scriptsize  0}
\put(69,85){\scriptsize  1}
\put(- 9, 9){\scriptsize   -1}
\put(-16,19){\scriptsize -2/3}
\put(-16,29){\scriptsize -1/3}
\put(- 8,39){\scriptsize    0}
\put(-14,49){\scriptsize  1/3}
\put(-14,59){\scriptsize  2/3}
\put(- 8,69){\scriptsize    1}
\put( 85, 9){\scriptsize   -1}
\put( 85,19){\scriptsize -2/3}
\put( 85,29){\scriptsize -1/3}
\put( 85,39){\scriptsize    0}
\put( 85,49){\scriptsize  1/3}
\put( 85,59){\scriptsize  2/3}
\put( 85,69){\scriptsize    1}
\put( 17,-9){\vector(1,0){15}}
\put( 17,85){\vector(1,0){15}}
\put(-20,15){\vector(0,1){15}}
\put(100,15){\vector(0,1){15}}
\put( 22,-8){\scriptsize$\delta  $}
\put( 22,86){\scriptsize$\delta  $}
\put(-24,20){\scriptsize$\epsilon$}
\put(101,20){\scriptsize$\epsilon$}
\put(40,45){\scriptsize (1)}
\put(36,55){\scriptsize (2)}
\put(39,25){\scriptsize (3)}
\put(18,43){\scriptsize (4)}
\put(57,36){\scriptsize (4)}
\put( 2,43){\scriptsize (5)}
\put(70,34){\scriptsize (5)}
\put(16,66){\scriptsize (6)}
\put(57,60){\scriptsize (6)}
\put(15,18){\scriptsize (6)}
\put(58,12){\scriptsize (6)}
\put(40,45){\vector(0,-1){4}}
\end{picture}}
\hspace*{2cm}\parbox[b]{45mm}{\caption{Domain of $\epsilon,\delta$ 
determining the shape of the gyro \label{fig}}}
\end{figure}
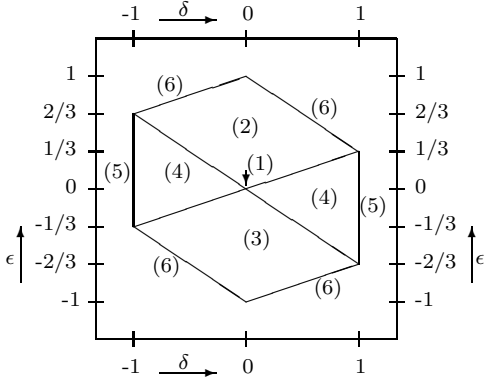
\begin{itemize}

\item[(1)] $\epsilon=0,\,\delta= 0$: \\
	   This is the pure sphere of radius $1$. 

\item[(2)] $0<|\epsilon| <1\,,\; \tfrac{1}{6} \epsilon {+} \tfrac{1}{2} 
|\epsilon| \leq \delta <  1 {+} \tfrac{1}{6} \epsilon 
{-} \tfrac{1}{2} |\epsilon|$: \\
The southern part of the sphere is rotary-grinded to a paraboloid.

\item[(3)] $0\leq|\epsilon| <1\,,\; {-} 1 {+}\tfrac{1}{6} \epsilon 
{+} \tfrac{1}{2} |\epsilon| < \delta \leq \tfrac{1}{6} \epsilon 
{-} \tfrac{1}{2} |\epsilon|$: \\
The northern part of the sphere is rotary-grinded to a paraboloid.

\item[(4)] $0<|\epsilon| <1\,,\; \tfrac{1}{6} \epsilon 
{-} \tfrac{1}{2} |\epsilon| < \delta < \tfrac{1}{6} \epsilon 
{+} \tfrac{1}{2} |\epsilon|$: \\
Both the northern and southern parts of the sphere are rotary-grinded to 
paraboloids but some region of the sphere remains. An example of this situation 
is shown in figure \ref{gyro}.

\item[(5)] $|\epsilon|=1\,,\; \delta \neq {\pm} 1 {+}\tfrac{1}{6} \epsilon 
{\mp} \tfrac{1}{2} |\epsilon|$: \\
The sphere is rotary-grinded to a lense composed of two paraboloids.

\item[(6)] $-1 \leq \epsilon \leq 1\,,\; \delta = {\pm} 1 {+} \tfrac{1}{6} 
\epsilon {\mp} \tfrac{1}{2} |\epsilon|$: \\
The sphere is rotary-grinded to a string of length $1$.

\item[$\bullet~$] $|\delta{-}\tfrac{1}{6} \epsilon| \leq \tfrac{1}{2}
{-}\tfrac{1}{2}|\epsilon|$: \\
The sphere's equator belongs to the gyro so that the maximal radius is $1$.

\item[$\bullet~$] $1 {-}\tfrac{1}{2}|\epsilon| \geq |\delta{-}\tfrac{1}{6} 
\epsilon| \geq \tfrac{1}{2}{-}\tfrac{1}{2}|\epsilon|$: \\
One paraboloid passes the sphere's equator and the maximal 
radius is $\sqrt{1-2(\tfrac{1}{2}{-}\tfrac{1}{2}|\epsilon|{-} 
|\delta{-}\tfrac{1}{6} \epsilon|)^2}$.

\item[$\bullet~$] The height of the gyro is 
$2{-}(|\delta{-}\tfrac{1}{6} \epsilon| {+} \tfrac{1}{2} |\epsilon|)$ in cases 
(1),(2),(3),(6) and $2{-}|\epsilon|$ in cases (1),(4),(5),(6).

\end{itemize}

We now derive the metric properties of these gyros. The 
first step is to compute $\|[Y,\pi(\boldsymbol{g},\boldsymbol{a},e)]\|$, which 
we demand to be bounded by $1$. The calculation splits into that for 
leptons $(\ell)$ and quarks $(q)$, see \eqref{pi} and \eqref{Y}. We confine 
our attention to the quark sector. It is convenient to use the $C^*$-property 
and to evaluate
\begin{align}
&[Y_q, \pi_q(\boldsymbol{g},\boldsymbol{a},e)]^* 
[Y_q,\pi_q(\boldsymbol{g},\boldsymbol{a},e)] 
= \mbox{\small$\left( \begin{array}{cc} L & 0 \\ 0 & R \end{array} \right)$}\;,
\\ 
&L = \!\mbox{\small${\renewcommand{\arraycolsep}{0pt} 
\left( \begin{array}{cc} 
\begin{array}{c} (a{-}e)^2 \one_3 \otimes \mathcal{M}_u \mathcal{M}_u^* \\
{+} (b^2{+}c^2) \one_3 \otimes \mathcal{M}_d \mathcal{M}_d^* \end{array} &
\!\!\!(a{-}e)(b {-} \mathrm{i} c) \one_3 \otimes (\mathcal{M}_u \mathcal{M}_u^* 
{-} \mathcal{M}_d \mathcal{M}_d^*) 
\\
(a{-}e)(b {+} \mathrm{i} c) \one_3 \otimes (\mathcal{M}_u \mathcal{M}_u^* 
{-} \mathcal{M}_d \mathcal{M}_d^*) \!\!\!
& 
\begin{array}{c} (b^2{+}c^2) \one_3 \otimes \mathcal{M}_u \mathcal{M}_u^* \\
{+} (a{-}e)^2 \one_3 \otimes \mathcal{M}_d \mathcal{M}_d^* \end{array}
\end{array} \right) }$}, \notag
\\
& R= \mbox{\small$ \left( \begin{array}{cc} 
((a{-}e)^2{+}b^2{+}c^2) \one_3 \otimes \mathcal{M}_u^* \mathcal{M}_u & 0 \\
0 & ((a{-}e)^2{+}b^2{+}c^2) \one_3 \otimes \mathcal{M}_d^* \mathcal{M}_d 
\end{array} \right) $}\,. \notag
\end{align}
Unitary transformation $U L U^*$ of the left sector $L$, with 
\begin{align*}
U &= \mbox{\small$ \left( \begin{array}{cc} 
\cos \tau \one_3 \otimes \one_3 & 
\mathrm{e}^{\mathrm{i} \sigma} \sin \tau \one_3 \otimes \one_3 \\
{-}\sin \tau \one_3 \otimes \one_3 &
\mathrm{e}^{\mathrm{i} \sigma} \cos \tau \one_3 \otimes \one_3 \end{array} 
\right) $} \in \mathrm{U(2)} \one_3 \otimes \one_3 ~, \\
\cos \tau &= \dfrac{a{-}e}{\sqrt{(a{-}e)^2{+}b^2{+}c^2}}~,\quad 
\sin \tau = \dfrac{\sqrt{b^2{+}c^2}}{\sqrt{(a{-}e)^2{+}b^2{+}c^2}}~,\quad 
\mathrm{e}^{\mathrm{i} \sigma}= \dfrac{b{-}\mathrm{i} c}{\sqrt{b^2{+}c^2}}~,
\end{align*}
yields the matrix
\[
((a{-}e)^2{+}b^2{+}c^2) \,\mathrm{diag}(\one_3 {\otimes} 
\mathcal{M}_u \mathcal{M}_u^*\,,\,
\one_3 {\otimes} \mathcal{M}_d \mathcal{M}_d^*\,,\,\one_3 {\otimes} \
\mathcal{M}_u^* \mathcal{M}_u\,,\,
\one_3 {\otimes} \mathcal{M}_d^* \mathcal{M}_d)~.
\]
Thus, the eigenvalues of $[Y, \pi(\boldsymbol{g},\boldsymbol{a},e)]^* 
[Y,\pi(\boldsymbol{g},\boldsymbol{a},e)]$ are $((a-e)^2+b^2+c^2)$ times the 
squared fermion masses and, because the mass of the top quark $m_t$ is the 
largest one, we have
\begin{equation}
\|[Y,\pi(\boldsymbol{g},\boldsymbol{a},e)]\|=\sqrt{(a-e)^2+b^2+c^2}\,m_t~.
\label{Ypi}
\end{equation}

The second step is to compute 
\begin{align*}
\chi_{\boldsymbol{\epsilon},\boldsymbol{\alpha},\delta}(\boldsymbol{g},
\boldsymbol{a},e) &- \chi_{\boldsymbol{\epsilon}',\boldsymbol{\alpha}',
\delta'} (\boldsymbol{g},\boldsymbol{a},e) \\
 =(\alpha{-}\alpha') &(a{-}e) + (\beta{-}\beta') b + (\gamma{-}\gamma') c +
2 (\delta{-}\delta') e + {\textstyle \sum_{i=1}^8} 
(\epsilon^j{-}\epsilon^j{}') g^j~.
\end{align*}
For $\delta \neq \delta'$ take $(\boldsymbol{g},a,b,c,e)=
(\boldsymbol{0},a,0,0,a)$, which on one hand gives 
$\|[Y,\pi(\boldsymbol{0},a,0,0,a)]\|=0$ 
for all $a$ and on the other hand $|\chi_{\boldsymbol{\epsilon},
\boldsymbol{\alpha},\delta}(\boldsymbol{0},a,0,0,a)
-\chi_{\boldsymbol{\epsilon}',\boldsymbol{\alpha}', 
\delta'} (\boldsymbol{0},a,0,0,a)|= 2 |\delta{-}\delta'|\,|a|$, which is 
obviously unbounded. The same effect happens if there is 
$\epsilon^j \neq \epsilon^j{}'$ for at least one $j=1,\dots,8$. In other words, 
the distance between functionals with different 
$\{\boldsymbol{\epsilon},\delta\}$ is infinite. 

Let us thus compute the distance between functionals with fixed 
$\{\boldsymbol{\epsilon},\delta\}$:
\[
|\chi_{\boldsymbol{\epsilon},\boldsymbol{\alpha},\delta}
(\boldsymbol{g},\boldsymbol{a},e) 
-\chi_{\boldsymbol{\epsilon},\boldsymbol{\alpha}',\delta}
(\boldsymbol{g},\boldsymbol{a},e)| =
|(\alpha{-}\alpha') (a{-}e) + (\beta{-}\beta') b + (\gamma{-}\gamma') c|~. 
\]
Under the condition $\|[Y,\pi(\boldsymbol{g},\boldsymbol{a},e)]\|
=\sqrt{(a-e)^2+b^2+c^2}\, m_t \leq 1$, this number is bounded by 
\begin{equation}
\mathrm{dist}(\boldsymbol{\alpha},\boldsymbol{\alpha}') / m_t \equiv 
\sqrt{(\alpha{-}\alpha')^2 + (\beta{-}\beta')^2 + 
(\gamma{-}\gamma')^2} \,/ m_t~. 
\label{distm}
\end{equation}
This means that the space of functionals $\chi_{\boldsymbol{\epsilon},
\boldsymbol{\alpha},\delta}$, for fixed parameters $\{\boldsymbol{\epsilon},
\delta\}$, is just the Euclidean space $\mathbb{R}^3$ equipped with the usual 
Euclidean distance (scaled by $1/m_t$). If we restrict the functionals 
$\chi_{\boldsymbol{\epsilon},\boldsymbol{\alpha},\delta}$ and 
$\chi_{\boldsymbol{\epsilon},\boldsymbol{\alpha}',\delta}$ to be points on a 
gyro of functionals of norm 1, their distance is equal to the Euclidean length 
(in units of $1/m_t$) of the string through the interior that connects the 
points on the gyro. Gyros associated to different parameters 
$\{\boldsymbol{\epsilon},\delta\}$ are infinitely distant from each other. 
Thus, the geometry of the standard model matrix $L$-cycle is a nine-parametric 
family (the parameters are $\epsilon^j$ and $\delta$) of infinitely distant 
gyros. This picture has a natural physical interpretation. 
The three massive Yang--Mills fields $W^\pm$ and $Z$ yield in a first step 
$\mathbb{R}^3$ and the norm=1 requirement selects a certain hypersurface in 
$\mathbb{R}^3$ -- our gyro. The nine-dimensional 
disconnectedness reflects the nine massless Yang--Mills fields (photon and 
gluons) of the standard model. Again, the norm=1 condition selects a compact 
region of $\mathbb{R}^9$ as shown in figure \ref{fig}.

\section{The continuous part of the standard model}
\label{cont}

Now we add spacetime to investigate the metric structure of the continuous part 
of the standard model. We only consider functionals on 
\[
\mathfrak{g}=C^\infty(M) \otimes \big(\mathrm{su(3)} {\oplus} \mathrm{su(2)} 
{\oplus} \mathrm{u(1)}\big) \ni \big(\boldsymbol{g},\boldsymbol{a},e\big)
\]
which are of the form
\begin{equation}
\chi_{\boldsymbol{\epsilon},\boldsymbol{\alpha},\delta;p}
(\boldsymbol{g},\boldsymbol{a},e) := \alpha a(p) + \beta b(p) 
+ \gamma c(p) + (2 \delta {-} \alpha) e(p) + 
\mbox{\small$\displaystyle \sum_{j=1}^8$} \epsilon^j g^j(p)\,, \label{eadp} 
\end{equation}
with $\{\boldsymbol{\epsilon},\boldsymbol{\alpha},\delta\}$ fixed for all 
functionals under consideration. Here, $e$ is a function on $M$ and $e(p)$ 
its value at the point $p \in M$, and so on. Moreover, we evaluate the distance 
by means of the Dirac operator $D=\mathrm{i} \dslash$ of the spin connection. 
The Hilbert space is $\mathcal{H}=L^2(\mathcal{S}) \otimes \mathbb{C}^{48}$ and 
the representation $\pi: \mathfrak{g} \to \mathcal{B}(\mathcal{H})$ coincides 
pointwise with the matrix representation \eqref{pi}. 

We investigate the following problem: For given $p,q \in M$ find the 
supremum of 
\begin{align}
&|\chi_{\boldsymbol{\epsilon},\boldsymbol{\alpha},\delta;p}
(\boldsymbol{g},\boldsymbol{a},e)-
\chi_{\boldsymbol{\epsilon},\boldsymbol{\alpha},\delta;q}
(\boldsymbol{g},\boldsymbol{a},e)| \notag \\
&{}~~= \big| \alpha (a(p) {-}a(q))
+ \beta (b(p){-}b(q)) + \gamma (c(p){-}c(q)) \label{xx} \\
&{}~~ + (2 \delta {-}\alpha) (e(p){-}e(q)) 
+ {\textstyle \sum_{j=1}^8 \epsilon^j (g^j(p){-}g^j(q))} \big| \notag
\end{align}
under the condition [see \eqref{Df0}]
\begin{equation}
\|[\mathrm{i} \dslash,\pi(\boldsymbol{g},\boldsymbol{a},e)]\| 
=\|\pi\big(\gamma(d\boldsymbol{g}),\gamma(d\boldsymbol{a}),\gamma(de)\big)\| 
\leq 1~.
\label{idpi}
\end{equation}

In the same way as in section \ref{commut} we can replace in \eqref{idpi} 
partial differentiations by differentiations along curves. Note that each of 
the 12 real functions parametrizing $\mathfrak{g}$ is differentiated 
independently, and the supremum is found by optimization of these 12 curves. 
Differentiation along a common curve yields a smaller value than the norm, 
so that \eqref{idpi} implies
\begin{equation}
\sup_{\mathcal{C}} \left|\!\left| \gamma(ds) \otimes 
\pi\Big( \frac{d\boldsymbol{g}}{ds},\frac{d\boldsymbol{a}}{ds},
\frac{de}{ds}\Big) \right|\!\right| \leq z \leq 1~.
\end{equation}
If $s$ is the arc length we have $(\gamma(ds))^2=\one_4$, see \eqref{Df}, so 
that $\gamma(ds)$ can be diagonalized to a matrix containing $\pm 1$ on 
the diagonal. This yields with \eqref{max}
{\allowdisplaybreaks[1]
\begin{align*} 
z &\geq \sup_{\mathcal{C}} \left|\!\left| \gamma(ds) \otimes 
\pi\Big( \frac{d\boldsymbol{g}}{ds},\frac{d\boldsymbol{a}}{ds},
\frac{de}{ds}\Big) \right|\!\right|   
\\
&= \sup_{\mathcal{C},x} \max\Big( \left|\frac{de}{ds}\right| 
\!{+}\! \left|\!\left| \frac{d\boldsymbol{a}}{ds}\right|\!\right| ,\, 
2\left|\frac{de}{ds}\right| ,\, 
\left|\frac{1}{3} \frac{de}{ds} {+} \frac{dg_i}{ds}\right| 
\!{+}\! \left|\!\left| \frac{d\boldsymbol{a}}{ds}\right|\!\right| ,\, 
\left|\frac{1}{3} \frac{de}{ds} {+} \frac{dg_i}{ds}\right| 
\!{+}\! \left|\frac{de}{ds} \right| \Big). 
\end{align*}}

As in section \ref{commut} we can replace the optimized differentiation by an 
optimized difference quotient, see the steps from \eqref{Df} to \eqref{Df1}.
The result is
\begin{align*} 
z &\geq \sup_{p \neq q} \Big\{ \mbox{\small$\dfrac{1}{\mathrm{dist}(p,q)}$} \:
\max \big( \begin{array}[t]{l} 
|e(p){-}e(q)|{+} \|\boldsymbol{a}(p){-}\boldsymbol{a}(q)\|\,,\, 
2|e(p){-}e(q)|\,,\, \\[1ex]
|(\tfrac{1}{3} e {+} g_i)(p){-}(\tfrac{1}{3} e {+} g_i)(q)| 
{+} \|\boldsymbol{a}(p){-}\boldsymbol{a}(q)\| \,,\, \\[1ex]
|(\tfrac{1}{3} e {+} g_i)(p){-}(\tfrac{1}{3} e {+} g_i)(q)| {+} |e(p){-}e(q)|~ 
\big) \Big\}~, \end{array} 
\end{align*}
which gives for any $p,q \in M$, $p\neq q$, the inequality
\begin{equation}
\mbox{\small$\dfrac{1}{\mathrm{dist}(p,q)}$} \:
\max \big( \begin{array}[t]{l} 
|e(p){-}e(q)|{+} \|\boldsymbol{a}(p){-}\boldsymbol{a}(q)\|\,,\, 
2|e(p){-}e(q)|\,,\, \\[1ex]
|(\tfrac{1}{3} e {+} g_i)(p){-}(\tfrac{1}{3} e {+} g_i)(q)| 
{+} \|\boldsymbol{a}(p){-}\boldsymbol{a}(q)\| \,,\, \\[1ex]
|(\tfrac{1}{3} e {+} g_i)(p){-}(\tfrac{1}{3} e {+} g_i)(q)| {+} |e(p){-}e(q)|~ 
\big) \leq \tilde{z}\;, \end{array} 
\label{maxd}
\end{equation}
with $0 < \tilde{z} \leq z \leq 1$. Comparison with our previous problem (to 
find $\max |F|$) in section \ref{matrix} suggests the replacements
{\allowdisplaybreaks[1]
\begin{gather*}
\boldsymbol{a}(p){-}\boldsymbol{a}(q) \mapsto \hat{\boldsymbol{a}}\, 
\mathrm{dist}(p,q)\,,\qquad e(p){-}e(q) \mapsto \hat{e}\, 
\mathrm{dist}(p,q)\,,\, \\
(\tfrac{1}{3} e {+} g_i)(p) - (\tfrac{1}{3} e {+} g_i)(q) \mapsto 
(\tfrac{1}{3} \hat{e} {+} \hat{g}_i )\, \mathrm{dist}(p,q)\,,
\end{gather*}}
where $\hat{\boldsymbol{a}} \in \mathrm{su(2)}$, $\hat{e} \in \mathbb{R}$, 
and $\mathrm{i}\hat{g}_i \in \mathrm{i}\mathbb{R}$ are the eigenvalues of 
$\hat{\boldsymbol{g}} \in \mathrm{su(3)}$. Then, our problem 
\eqref{xx}--\eqref{idpi} reduces to the matrix problem
\[
\mathrm{dist}(\chi_{\boldsymbol{\epsilon},\boldsymbol{\alpha},\delta;p},
\chi_{\boldsymbol{\epsilon},\boldsymbol{\alpha},\delta;q}) = 
\sup_{\hat{\boldsymbol{g}},\hat{\boldsymbol{a}},\hat{e}}  
\big\{ \mathrm{dist}(p,q) \, |\chi_{\boldsymbol{\epsilon},
\boldsymbol{\alpha},\delta} (\hat{\boldsymbol{g}},\hat{\boldsymbol{a}},
\hat{e})|\,:~ \|\pi(\hat{\boldsymbol{g}},\hat{\boldsymbol{a}},\hat{e})\| 
\leq \tilde{z}\, \big\} \,.
\]
But this means nothing else than
\begin{equation}
\mathrm{dist}(\chi_{\boldsymbol{\epsilon},\boldsymbol{\alpha},\delta;p},
\chi_{\boldsymbol{\epsilon},\boldsymbol{\alpha},\delta;q}) 
= \tilde{z} \,\mathrm{dist}(p,q) \,\|\chi_{\boldsymbol{\epsilon},
\boldsymbol{\alpha},\delta} \| = \tilde{z}\,\mathrm{dist}(p,q) 
\leq \,\mathrm{dist}(p,q)~,
\label{distc}
\end{equation}
because the functionals $\chi_{\boldsymbol{\epsilon},
\boldsymbol{\alpha},\delta}$ satisfy 
$\|\chi_{\boldsymbol{\epsilon},\boldsymbol{\alpha},\delta} \|=1$ if 
$\{\boldsymbol{\epsilon},\boldsymbol{\alpha},\delta\}$ determine a gyro. 

We will now prove that \eqref{distc} is actually an equality. For this purpose 
we consider optimized matrices multiplied by the distance function:
\begin{gather}
\boldsymbol{a}_p(x)=\mathrm{dist}(p,x)\, \hat{\boldsymbol{a}}\,, \quad
\boldsymbol{g}_p(x)=\mathrm{dist}(p,x)\, \hat{\boldsymbol{g}}\,, \quad
e_p(x)=\mathrm{dist}(p,x) \,\hat{e}\,, 
\label{opt} \\
\|\pi(\hat{\boldsymbol{g}},\hat{\boldsymbol{a}},\hat{e})\|=1\,,\quad
|\alpha \hat{a} + \beta \hat{b} + \gamma \hat{c} + (2 \delta{-}\alpha) \hat{e}
+ {\textstyle \sum_{j=1}^8 \epsilon^j \hat{g}^j}| =1\,. 
\notag
\end{gather}
This gives for \eqref{idpi}
\[
\|[\mathrm{i} \dslash,\pi(\boldsymbol{g}_p,\boldsymbol{a}_p,e_p)]\| = 
\sup_x |\dslash(\mathrm{dist}(p,x))|\, 
\|\pi(\hat{\boldsymbol{g}},\hat{\boldsymbol{a}},\hat{e})\|=1
\]
due to \eqref{Df1} on one hand and on the other hand for \eqref{xx}
\begin{align}
&|\chi_{\boldsymbol{\epsilon},\boldsymbol{\alpha},\delta;p}
(\boldsymbol{g}_p,\boldsymbol{a}_p,e_p)-
\chi_{\boldsymbol{\epsilon},\boldsymbol{\alpha},\delta;q}
(\boldsymbol{g}_p,\boldsymbol{a}_p,e_p)| \notag \\
&{}~~= \mathrm{dist}(p,q) \,\big| \alpha \hat{a} + \beta \hat{b} 
+ \gamma \hat{c} + (2 \delta {-} \alpha) \hat{e} 
+ {\textstyle \sum_{j=1}^8 \epsilon^j \hat{g}^j} \big| =\mathrm{dist}(p,q)~.
\label{diste}
\end{align}
Hence, we get the nice result that the distance between functionals 
$\chi_{\boldsymbol{\epsilon},\boldsymbol{\alpha},\delta;p}$ 
($\boldsymbol{\epsilon},\boldsymbol{\alpha},\delta$ fixed and determining a 
gyro, $p$ variable) is equal to the geodesic distance between 
the points $p \in M$. However, this result is preliminary because the standard 
model Dirac operator is not $\mathrm{i} \dslash$ (as we used in this section) 
but the Dirac--Yukawa operator that involves the fermionic mass matrix $Y$, see 
\eqref{Y}.

\section{The full standard model}
\label{full}

Here we unite matrix part and continuous part to the full standard 
model. We compute the distance between functionals \eqref{eadp}, where we now
permit a variation of $\{\boldsymbol{\epsilon},\boldsymbol{\alpha}, 
\delta\}$. Moreover, we evaluate the metric with the full Dirac-Yukawa operator 
$D=\mathrm{i} \dslash+ \gamma^5 Y$, which is the sum of the previous cases. 
Lie algebra and Hilbert space are as in section \ref{cont}. We have to find 
the supremum of 
\begin{equation}
\begin{split}
|\chi_{\boldsymbol{\epsilon}, \boldsymbol{\alpha},\delta;p} &
(\boldsymbol{g},\boldsymbol{a},e)-
\chi_{\boldsymbol{\epsilon}',\boldsymbol{\alpha}',\delta';q}
(\boldsymbol{g},\boldsymbol{a},e)|  \\
=& \big| \alpha (a(p) {-}a(q))
+ \beta (b(p){-}b(q)) + \gamma (c(p){-}c(q)) \\
+& (2 \delta {-}\alpha) (e(p){-}e(q)) 
+ {\textstyle \sum_{j=1}^8 \epsilon^j (g^j(p){-}g^j(q))}  \\
+& (\alpha{-}\alpha') (a(q){-}e(q)) + (\beta{-}\beta') b(q) 
+ (\gamma{-}\gamma') c(q)  \\
+& 2 (\delta {-}\delta') e(q) 
+ {\textstyle \sum_{j=1}^8 (\epsilon^j{-}\epsilon^j{}') g^j(q)} \big|
\end{split}
\label{xxf}
\end{equation}
under the condition
\begin{equation}
\|u+v\| \leq 1~,\qquad 
u:=[\mathrm{i} \dslash,\pi(\boldsymbol{g},\boldsymbol{a},e)]~,\quad
v:=\gamma^5 [Y,\pi(\boldsymbol{g},\boldsymbol{a},e)]~.
\label{cf}
\end{equation}

The exact solution of the problem \eqref{xxf}--\eqref{cf} would involve the 
diagonalization of $4\times 4$-matrices, which is too ambitious. We therefore 
will give an exact lower bound for the distance and estimate an upper bound.
The lower bound is found by investigation of the extremal cases $u=0$ or $v=0$. 
The case $u=0$ is achieved by taking constant matrices and leads back to 
section \ref{matrix}. It is clear that the distance is infinity unless we 
require $\delta=\delta'$ and $\boldsymbol{\epsilon}=\boldsymbol{\epsilon}'$.
Under this condition, the result \eqref{distm} holds and we get
\begin{equation}
\mathrm{dist}(\chi_{\boldsymbol{\epsilon},\boldsymbol{\alpha},\delta;p},
\chi_{\boldsymbol{\epsilon},\boldsymbol{\alpha}',\delta;q})
\geq \mathrm{dist}(\boldsymbol{\alpha},\boldsymbol{\alpha}')\,/\,m_t~.
\end{equation}

We now adjust $v=0$, which means $a=e$ and $b=c=0$. The lesson of section 
\ref{cont} was that the optimum is attained in the class ($\mathrm{dist}(p,x)$ 
times an appropriate matrix) of elements of $\mathfrak{g}$. This class 
corresponds to the first line of \eqref{opt}, with $\hat{a}=\hat{e}$ and 
$\hat{b}=\hat{c}=0$. From \eqref{diste} we conclude
\[
\mathrm{dist}(\chi_{\boldsymbol{\epsilon},\boldsymbol{\alpha},\delta;p},
\chi_{\boldsymbol{\epsilon},\boldsymbol{\alpha}',\delta;q})
\geq \mathrm{dist}(p,q) \: \sup_{\hat{e},\hat{\boldsymbol{g}}} \, 
\dfrac{|2 \delta \hat{e} {+} \epsilon \hat{g}_3|}{ 
\|\pi(\hat{\boldsymbol{g}},\hat{e},\hat{e})\|}~.
\]
The search for the extrema of $\hat{F}=(2 \delta \hat{e} {+} \epsilon 
\hat{g}_3)/\pi(\hat{\boldsymbol{g}},\hat{e},\hat{e})$ is easier than the 
problem solved in section \ref{matrix}. From \eqref{pi} we see that the left 
and right norms are identical and equal to the right norm in \eqref{lrc}. The 
table corresponding to table \ref{table} reads:
\[
\mbox{\small$
\renewcommand{\arraystretch}{1.9}
\begin{array}{rcl}
\mathrm{N}^{\underline{\mathrm{o}}} & \hat{e} & \hat{F} \\ \hline
1  & -\tfrac{3}{2} |\hat{g}_1| & 
\dfrac{ -3 \delta |\hat{g}_1| + \epsilon \hat{g}_3}{3 \hat{g}_1} 
\\
2  & -\tfrac{3}{2} (\hat{g}_3{-}|\hat{g}_1|) & 
\dfrac{ 3 \delta |\hat{g}_1| + (\epsilon{-} 3\delta) \hat{g}_3}{
2 \hat{g}_3{-}|\hat{g}_1|} 
\\
3 & 0 & \dfrac{\epsilon \hat{g}_3}{\hat{g}_3} 
\\
4  & \tfrac{3}{2} \hat{g}_3 & 
\dfrac{(\epsilon{+} 3\delta) \hat{g}_3}{3 \hat{g}_3} 
\\
5 & \pm M & \dfrac{\pm 2 M \delta}{2 M} 
\end{array}$}
\]
This yields 
{\allowdisplaybreaks[1]
\begin{gather}
\|\hat{F}\|=B=\max(|\epsilon|,\,|\delta{-}\tfrac{1}{6} \epsilon|{+}
\tfrac{1}{2} |\epsilon|) \leq 1 \notag 
\\
\Rightarrow \quad \mathrm{dist}(\chi_{\boldsymbol{\epsilon},
\boldsymbol{\alpha},\delta;p}, \chi_{\boldsymbol{\epsilon},
\boldsymbol{\alpha}',\delta;q}) \geq B\,\mathrm{dist}(p,q) ~.
\end{gather}}
Thus, $\mathrm{dist}(\chi_{\boldsymbol{\epsilon},\boldsymbol{\alpha},\delta;p}, 
\chi_{\boldsymbol{\epsilon},\boldsymbol{\alpha}',\delta;q})$ is bounded up to 
the scale factor $B$ by the spacetime distance $\mathrm{dist}(p,q)$, where $B$ 
becomes $1$ only on the boundary of the parameter region 
$\{\epsilon,\delta\}$ of allowed gyros (figure \ref{fig}).

If we now rise $|a-e|,|b|,|c|$, then the distance will also grow at first due 
to the part $\alpha(a{-}e)(p) {-}\alpha'(a{-}e)(q) + \beta b(p){-}\beta'b(q) 
+ \gamma c(p){-}\gamma'c(q)$ in \eqref{xxf}. But very soon this growth is 
compensated by the necessity to decrease $u$ at expense of the growth of $v$. 
We have 
\begin{multline*}
|\alpha(a{-}e)(p) {-}\alpha'(a{-}e)(q) + \beta b(p){-}\beta'b(q) 
+ \gamma c(p){-}\gamma'c(q)| \\
\leq 2 \,\sup_x \sqrt{(a{-}e)^2+b^2+c^2} \leq 2/m_t
\end{multline*}
for $\|v\| =\|\gamma^5 v \|\leq 1$, see \eqref{Ypi}. This means
\[
\mathrm{dist}(\chi_{\boldsymbol{\epsilon},\boldsymbol{\alpha},\delta;p},
\chi_{\boldsymbol{\epsilon},\boldsymbol{\alpha}',\delta;q}) \leq 
B\, \mathrm{dist}(p,q) + 2/m_t~,
\]
and we find the final result
\begin{multline}
\max\big\{ B\,\mathrm{dist}(p,q)\,,\, \mathrm{dist}(\boldsymbol{\alpha},
\boldsymbol{\alpha}')/ m_t \big\} \\
\leq \mathrm{dist}(\chi_{\boldsymbol{\epsilon},\boldsymbol{\alpha},
\delta;p}, \chi_{\boldsymbol{\epsilon},\boldsymbol{\alpha}',\delta;q}) 
\leq B\,\mathrm{dist}(p,q) + 2/m_t~.
\end{multline}
The precise value of $\mathrm{dist}(\chi_{\boldsymbol{\epsilon},
\boldsymbol{\alpha},\delta;p}, \chi_{\boldsymbol{\epsilon},
\boldsymbol{\alpha}',\delta;q})$ is not so important, its boundedness suffices 
for a physical discussion.

\section{Physical interpretation}

Note that $\mathrm{dist}(\boldsymbol{\alpha},\boldsymbol{\alpha}')/m_t \leq 
2 /m_t \approx 2.3 \,\cdot\, 10^{-16}\,\mathrm{cm}$. No measurement device for 
macroscopic distances has a precision of $10^{-16}\,\mathrm{cm}$. Hence, 
for geodesic distances $\mathrm{dist}(p,q)$ of atomic size or larger, the 
geometry of the standard model is in accurate agreement with $B$ times the 
Riemannian geometry of the underlying manifold. At scales of the order of the 
inverse top quark mass however, corresponding to energies of the order 
$100\,\mathrm{GeV}$, spacetime should reveal a completely different structure. 
That what macroscopically is a point becomes an extended object~-- a gyro.

As we have seen, there is a nine-parametric family of infinitely distant 
worlds (or universes) whose points (on macroscopic scales) are gyros 
(on scales $1/m_t$). The scale factor $B$ is constant on each 
world. At first glance, this unobserved scale factor $B$ seems to favour the 
conclusion that in our world values for ${\boldsymbol{\epsilon},\delta}$ are 
realized which are on the boundary of the allowed values (figure \ref{fig}). 
This means that the gyros would be degenerated to strings or lenses. However, 
we should remember that we cannot see the ``actual'' spacetime 
manifold $M$. All our measurements are only able to detect the ``derived'' 
geometry, which is $B$ times the true one. And as the actual manifold is not 
relevant, there is no problem in saying that the true geometry is $(1/B)$ 
times the measured geometry. This means that any of the allowed values 
for $\{\boldsymbol{\epsilon},\delta\}$ according to figure \ref{fig} (except 
$\epsilon{=}\delta{=}0$) could be realized in our universe. The laws of physics 
should be the same on each world except for effects due to different scale 
factors $B$, which certainly lead to different ``constants'' of nature. 

This picture of the geometry of our world is probably not ultimate knowledge. 
The standard model is in accurate agreement with experiment only because 
today's experiments have a maximal resolution of the order 
$10^{-16}\,\mathrm{cm}$. At this resolution, spacetime consists of gyros. 
But this does not exclude the possibility that at higher 
resolutions the gyros show a fine structure in the sense that 
each of its points is a higher dimensional object itself. Grand unified 
theories for example contain further massive Yang--Mills fields and a plenty of 
additional Higgs fields. We therefore expect that further dimensions become 
apparent at GUT scales $(10^{15} \dots 10^{16}\,\mathrm{GeV})$. 

In other words, we recover the old Kaluza--Klein idea \cite{ka,kl} of 
additional spacetime dimensions, which are compactified to very small size so 
that they are not apparent in every day's life. The attempt to deduce the 
fundamental interactions from higher dimensional Riemannian geometry has a 
long history \cite{ka,kl,eb,ke,m}. But this approach has a severe 
shortcoming. One has to make a guess for the higher dimensional spacetime and 
then to reduce dimensions in order to obtain an effective theory in four 
dimensions. Although Manton has already found \cite{m} the six-dimensional 
geometry of the Salam-Weinberg model (which in some sense coincides with our 
result), this trial-and-error method was not very effective after all. We 
simply took the other direction: We started from the experimentally 
well-confirmed standard model (Lie algebra, fermionic Hilbert space, fermionic 
mass matrix) and computed directly the corresponding small scale geometry. The 
essential progress (apart from its effectiveness) of our method lies in the 
fact that it implements chiral fermions from the very beginning~-- an
obstacle for traditional Kaluza--Klein theories. Moreover, we obtain a 
geometric interpretation of the unbroken symmetries: They parametrize the 
copies of the world.

Now the question arises: What was wrong with previous Kaluza--Klein theories? 
--- The physical interpretation! One has mostly attempted to identify the 
additional dimensions with Yang--Mills fields. This is correct in so far as the 
gyros are hypersurfaces in $\mathbb{R}^3$, because the standard model contains 
three massive Yang--Mills fields. A different explanation is that the gyros are 
deformations of the $2$-sphere $S^2 \cong \mathrm{SU(2)}/\mathrm{U(1)}$, which 
could be related to the spontaneously broken symmetry group. But the size and 
shape of the gyros are fixed, there is no geometry other than dimension related 
to Yang--Mills fields. The geometry of the gyros is rather related to the Higgs 
field. This becomes apparent if one adopts ideas of the Chamseddine--Connes 
approach of noncommutative geometry \cite{cc}. There, one studies the 
spectral geometry of the full Dirac operator $D_A$ which includes Yang--Mills 
fields and Higgs fields. Let us drop the Yang--Mills fields. Any noncommutative 
formulation of the standard model tells us that the Dirac--Yukawa--Higgs 
operator is obtained from the Dirac--Yukawa operator by replacing all fermion 
masses $m_i$ by $\phi m_i$, where $\phi$ is the Higgs field whose vacuum 
expectation value $\langle \phi \rangle_0$ equals $1$. Now, the distance 
scale on the gyro becomes $1/(\phi m_t)$ instead of $1/m_t$, and is therefore 
subject to change if the Higgs field varies.  

Thus, we handle the gyros on the same footing as Riemannian 
spaces $M$. Introducing coordinates $x=\{x^0,x^1,x^2,x^3\}$ on $M$, the 
distance between points $x,x'$ is not the Euclidean distance $\|x-x'\|$ but 
obtained on infinitesimal level by taking the metric tensor $g_{\mu\nu}$ into 
consideration, $(ds)^2=g_{\mu\nu} dx^\mu dx^\nu$. Just as the Higgs field on 
the rigid gyro, the metric tensor determines the scale on the 
rigid coordinate space. The analogy between metric tensor and Higgs field as 
the scale on coordinate space goes further: Both have non-vanishing vacuum 
expectation value: $\langle g_{\mu\nu}\rangle_0=\eta_{\mu\nu}$ 
(or $\delta_{\mu\nu}$ in Euclidean framework; $\eta_{\mu\nu}$ is the Minkowski 
tensor) and $\langle \phi \rangle_0=1$. The diameter of the gyro is determined 
by the inverse top quark mass or, equivalently \cite{rw3}, by the inverse mass 
of the Higgs boson (see also \cite{m}). Our analogy then implies that the 
diameter of the coordinate space of the four dimensional manifold should be of 
the order of the inverse mass of the graviton, and therefore equal to infinity. 
This explains why four coordinates are expanded whereas the internal 
coordinates are compactified. Einstein told us \cite{e1,e2} that masses 
determine the geometry of spacetime~-- on large scales. Our result is the 
inverse: the small scale structure of spacetime (gyros) tells us that there 
exist massive particles in the universe. Isn't this a beautiful interplay 
between large and small scales? The small scale structure of spacetime 
generates the masses which in turn generate the large scale structure.

We live on one specific world of the nine-parametric family. We cannot 
establish any contact with the other universes, we nevertheless know of their 
existence: the nine massless Yang--Mills fields occurring in our (and any 
other) world are the carriers of this global information. All gyros of our 
world have the same shape, denoted by $\Sigma_{\delta,\epsilon}$. The 
information about this shape (better: of the object that replaces the gyro at 
GUT-energies) and about the four-dimensionality of the underlying manifold must 
have been present as early as the big bang. As already stressed, the 
\emph{shape} $\mathbb{R}^4 \times \Sigma_{\delta,\epsilon}$ (or $S^4 \times 
\Sigma_{\delta,\epsilon}\,$?) is fixed forever, that what evolves is the 
\emph{scale}. Big bang singularity means that the distance between any points 
on $\mathbb{R}^4 \times \Sigma_{\delta,\epsilon}$ becomes zero, because 
$g^{\mu\nu}$ and $\phi$ diverge. This behavior is very similar to a correlation 
length that diverges at a critical point.

\end{document}